\documentstyle[11pt]{article}

\newcommand{\be}{\begin{equation}}
\newcommand{\ee}{\end{equation}}
\newcommand{\bea}{\begin{eqnarray}}
\newcommand{\eea}{\end{eqnarray}}
\def\non{\nonumber}
\begin{document}
\title{ Classical Communication Cost of Quantum Teleportation}
\author{
  Chirag N Dekate
  \thanks{E-mail address : cdekat1@lsu.edu }\\
  Louisiana State University\\
  Baton Rouge, LA 70803\\
  USA
  }
\date{\today}
\maketitle
\subsection*{\centering Abstract}
{Quantum teleportation schemes in which operations are performed before establishing the quantum channel are not constrained by resource limits set in H.K.Lo \cite{CCCIQIP} and Bennett et al. \cite{rpqs}. We compare the standard teleportation protocol to the one proposed by Kak on the basis of the classical communication cost. Due to its unique architecture we study the problems in implementing Kak teleportation protocol.}


\subsection*{Introduction}
Quantum state teleportation and dense coding are the elementary components of quantum information science. Both use entanglement as a resource in achieving information processing capabilities far superior to classical communication and computing. In dense coding, the classical communication capacity of the quantum channel is doubled as one qubit is used to transmit two classical bits. H.K.Lo \cite{CCCIQIP} and Bennett et al. \cite{rpqs} discuss the optimal resource limit in teleportation and both conclude that teleportation of a quantum state needs at least 2 classical bits and 1 shared EPR resource. Teleportation protocols recently proposed by Kak \cite{KQTP} teleport achieve similar results with lesser classical communication cost. 
\\
\\
The second protocol described in Kak's paper uses 3 entangled qubits and 1.5 classical bits to achieve teleportation of an unknown state. Although the classical communication cost is less than the standard case, we leave this protocol out of our discussion because for every unknown qubit to be teleported three entangled qubits are required. 
\\ 
\\
This paper is arranged as follows: We first review the teleportation protocols proposed by Bennett, next we review the first teleportation protocol proposed by Kak. Then we discuss the implications of Kak teleportation protocol on classical communication cost and finally we look at the implementation issues when dealing with the Kak teleportation protocol.

\subsection*{Standard Quantum Teleportation Protocol (SQTP)} 
The SQTP \cite{SQTP} permits Alice to transmit an unknown qubit
$|\psi\rangle$  to Bob. The unknown qubit $|\psi\rangle$  can be written as 
$|\psi\rangle = \alpha|0\rangle + \beta|1\rangle $. Assuming that Alice and Bob possess
an entangled pair of qubits $|\phi\rangle = \frac{1}{\sqrt{2}}[|00\rangle + |11\rangle]$, then the initial 
state of all the 3 particles can be written as 

\vspace{0.2in}

$ \frac{1}{\sqrt{2}}[\alpha|000\rangle + \alpha|011\rangle + \beta|100\rangle + \beta|111\rangle]$

\vspace{0.2in}


\noindent
Alice sends her 2 qubits through a CNOT gate resulting in 

\vspace{0.2in}

$ \frac{1}{\sqrt{2}}[\alpha|000\rangle + \alpha|011\rangle + \beta|110\rangle + \beta|101\rangle]$

\vspace{0.2in}

\noindent
In the next step Alice sends her unknown qubit through a Hadamard Gate resulting in

\vspace{0.2in}

$ \frac{1}{2}[\alpha|000\rangle + \alpha|100\rangle + \alpha|011\rangle + \alpha|111\rangle + \beta|010\rangle - \beta|110\rangle + \beta|001\rangle - \beta|101\rangle]$

\vspace{0.2in}

\noindent
This can be rewritten as

\vspace{0.2in}

$ \frac{1}{2}[|00\rangle (\alpha|0\rangle + \beta|1\rangle) +
              |01\rangle (\alpha|1\rangle + \beta|0\rangle) + 
              |10\rangle (\alpha|0\rangle - \beta|1\rangle) +
              |11\rangle (\alpha|1\rangle - \beta|0\rangle) ] 
$               

\vspace{0.2in}

\begin{table}
 \begin{center}
  \begin{tabular}{|c|c|l|} \hline
    \multicolumn{1}{|c}{Alice's Result} &
    \multicolumn{1}{|c|}{Bob's Action (U)} \\ \hline \hline
 00 & No transform needed \\
 01 & X gate \\
 10 & Z gate \\
 11 & First Z then X gate \\ \hline
  \end{tabular}
  \caption{\textit{Standard protocol: Unitary transforms applied by Bob to recover unknown state}}
  \label{tb:SQTP}
 \end{center}
\end{table}

Alice measures the above state which collapses into one of the 4 classical 
states 00, 01, 10, 11. These two bits are sent to Bob over a classical channel
Since Bob must wait for the result from Alice, faster than light communication is
not possible. Bob can apply unitary transforms as shown in Table 1 to recover the
unknown state. The measurement of her qubits by Alice destroys
the initial state of her photon, thus preventing the violation of no-cloning 
theorem.

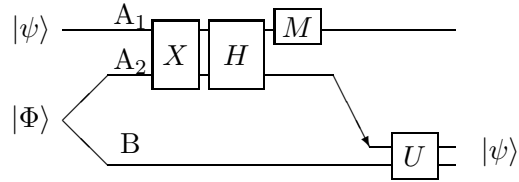
\bea
\setlength{\unitlength}{0.6mm}
\centering
\begin{picture}(90,53)
\put(-7,20){\makebox(10,10){$|\Phi\rangle$}}
\put(5,25){\line(1,1){10}}
\put(5,25){\line(1,-1){10}}
\put(15,43){\makebox(10,10){A$_1$}}
\put(15,33){\makebox(10,10){A$_2$}}
\put(15,15){\makebox(10,10){B}}
\put(-7,40){\makebox(10,10){$|\psi\rangle$}}
\put(15,35){\line(1,0){10}}
\put(15,15){\line(1,0){63}}
\put(25,32){\framebox(10,15){$X$}}
\put(35,35){\line(1,0){2.5}}
\put(37.5,32){\framebox(12,15){{$H$}}}
\put(49.5,45){\line(1,0){2.5}}
\put(49.5,35){\line(1,0){2.5}}
\put(52,42){\framebox(10,7.5){$M$}}
\put(52,35){\line(1,0){13}}
\put(62,45){\line(1,0){30}}
\put(65,35){\vector(1,-2){8}}
\put(73,19){\line(1,0){5}}
\put(78,12){\framebox(10,10){$U$}}
\put(88,19){\line(1,0){4}}
\put(88,15){\line(1,0){4}}
\put(97,14){\makebox(10,8){$|\psi\rangle$}}
\put(5,45){\line(1,0){20}}
\put(35,45){\line(1,0){2.5}}
\centering
\end{picture}
\non
\eea
{\bf SQTP}:\textit{In the above figure $|\psi\rangle$, $|\Phi\rangle$ correspond the unknown qubit and the entagled qubit respectively. The gates X, H, U represent XOR , Hadamard, Unitary transforms respectively. The symbol M represents Alice making the measurement.}  
\subsection*{Teleportation Protocols suggested by Kak} 

In Kak's first teleportation protocol\cite{KQTP} chained XOR transformations are applied as follows.
The initial state of all the 3 particles can be written as 

\vspace{0.2in}

$ \frac{1}{\sqrt{2}}[\alpha|000\rangle + \alpha|011\rangle + \beta|100\rangle + \beta|111\rangle]$

\vspace{0.2in}

\noindent
Alice sends the qubits through a chained XOR gate where in the first XOR
operates on the first 2 qubits and the second XOR gate operates on the 
last 2 qubits. After the first XOR we obtain:

\vspace{0.2in}

$ \frac{1}{\sqrt{2}}[\alpha|000\rangle + \alpha|011\rangle + \beta|110\rangle + \beta|101\rangle]$

\vspace{0.2in}

\noindent
This result is then passed through a second XOR to obtain:

\vspace{0.2in}

$ \frac{1}{\sqrt{2}}[\alpha|000\rangle + \alpha|010\rangle + \beta|111\rangle + \beta|101\rangle]$

\vspace{0.2in}

\noindent
The first and the third qubit are now fully entangled as a consequence of chained XOR transformations. The third particle is transferred to Bob. Applying the Hadamard transform on the first qubit we get:

\vspace{0.2in}

$ \frac{1}{2}[\alpha|000\rangle + \alpha|100\rangle + \alpha|010\rangle + \alpha|110\rangle + \beta|011\rangle - \beta|111\rangle + \beta|001\rangle - \beta|101\rangle]$

\vspace{0.2in}

\noindent
This can be rewritten as

\vspace{0.2in}

$ \frac{1}{2}[|00\rangle (\alpha|0\rangle + \beta|1\rangle) +
              |01\rangle (\alpha|0\rangle + \beta|1\rangle) + 
              |10\rangle (\alpha|0\rangle - \beta|1\rangle) +
              |11\rangle (\alpha|0\rangle - \beta|1\rangle) ] 
$               

\vspace{0.2in}

\noindent
This may be further simplified as 

\vspace{0.2in}

$ \frac{1}{2}[|0\rangle (|0\rangle + |1\rangle) (\alpha|0\rangle + \beta|1\rangle) +
              |1\rangle (|0\rangle + |1\rangle) (\alpha|0\rangle - \beta|1\rangle) 
             ] 
$               

\vspace{0.2in}

\begin{table}
 \begin{center}
  \begin{tabular}{|c|c|l|} \hline
    \multicolumn{1}{|c}{Alice's Result} &
    \multicolumn{1}{|c|}{Bob's Action} \\ \hline \hline
 0 & no transform needed \\
 1 & Z gate \\ \hline
  \end{tabular}
  \caption{\textit{Kak's protocol: Unitary transforms applied by Bob to recover unknown state}}
  \label{tb:Xname}
 \end{center}
\end{table}

\noindent
Alice now measures the first two qubits resulting in the collapse to 

\vspace{0.2in}

$ (\alpha|0\rangle + \beta|1\rangle) or (\alpha|0\rangle - \beta|1\rangle)               
$               

\vspace{0.2in}

\noindent
In this case only one classical bit (the result of measurement of first qubit) needs to be transmitted to
Bob. Bob can apply unitary transforms on the classical bit as shown in Table 2 to recover the unknown state.

\bea
\setlength{\unitlength}{0.6mm}
\centering
\begin{picture}(100,53)
\put(-7,20){\makebox(10,10){$|\Phi\rangle$}}
\put(5,15){\line(1,0){10}}
\put(5,35){\line(1,0){12}}
\put(5,45){\line(1,0){12}}
\put(6,43){\makebox(10,10){A$_1$}}
\put(6,34){\makebox(10,10){A$_2$}}
\put(6,13){\makebox(10,10){A$_3$}}
\put(-7,40){\makebox(10,10){$|\psi\rangle$}}
\put(15,15){\line(1,0){12}}
\put(17,33){\framebox(8,14){$X$}}
\put(24.5,35){\line(1,0){2.5}}
\put(25,45){\line(1,0){26.5}}
\put(27,13){\framebox(8,24){$X$}}
\put(35.5,15){\line(1,0){42}}
\put(35.5,35){\line(1,0){15.5}}
\put(60,35){\line(1,0){5}}
\put(88,15){\line(1,0){4}}
\put(51.5,32){\framebox(8,15){{$H$}}}
\put(59.5,45){\line(1,0){5}}
\put(38,43){\makebox(10,10){A$_1$}}
\put(38,34){\makebox(10,10){A$_2$}}
\put(38,13){\makebox(10,10){B$_1$}}
\put(65,42){\framebox(6,7.5){$M$}}
\put(71.5,45){\line(1,0){19}}
\put(65,35){\vector(1,-2){8}}
\put(73,19){\line(1,0){5}}
\put(78,12){\framebox(10,10){$U$}}
\put(88,19){\line(1,0){4}}
\put(97,14){\makebox(10,8){$|\psi\rangle$}}
\put(40,10){\vector(0,1){5}}
\put(35,0){\makebox(10,10){$T$}}
\centering
\end{picture}
\non
\eea
{\bf KQTP}:\textit{In the above figure $|\psi\rangle$, $|\Phi\rangle$ correspond the unknown qubit and the entagled qubit respectively. The gates X, H, U represent XOR , Hadamard, Unitary transforms respectively. The symbol M represents Alice making the measurement. T represents completion of the chained XOR and sharing of the entangled resource. Note the change from A$_3$ to B$_1$.}  

\subsection*{Discussion} 
Kak's teleportation protocol appears to be in violation of the requirements specified by H.K.Lo \cite{CCCIQIP} and Bennett et al. \cite{rpqs}. Both argue that the optimal resources for teleportation are 2 classical bits and 1 entangled qubit. This amount of classical resource corresponds to $2 \log_2 N$ bits for an N dimensional state. It can also be inferred that if a teleportation protocol uses less than $2 \log_2 N$ classical bits then the results of dense coding \cite{dc} would be violated. This argument however would not apply to all teleportation schemes. Protocols similar to the ones proposed by Kak will differ from the standard schemes in the way the quantum channel is prepared.
\\
\\
 Kak's protocol is more restrictive in terms of availability of the unknown state before the quantum resource is shared between Alice and Bob. However the protocol still conforms to the 2 stage standard starting with a pure unknown state and a shared EPR resource leading to the second stage where the unknown state is teleported to Bob. Teleportation of multiple qubits using Kak's protocol would require that for each unknown qubit to be teleported a corresponding unique quantum channel be established. Additionally teleportation of entangled states would not be possible under Kak's protocol as the chained XOR transformations would damage the state to be teleported.   
\\
\\
Remote state preparation \cite{rpqs} is another scheme of transmitting information which provides similar levels of optimal resource limits. In remote state preparation, Alice is given an infinite number of pure states out of which she needs to transmit only one to Bob. Since the distribution does not need to be random and due to Alice's prior knowledge of the pure state, remote state preparation protocols are different from teleportation. The classical communication cost in remote state preparation and Kak's teleportation protocol are same.  

\subsection*{Implementation Issues}
Practical implementations of teleportation have mainly focused on photonic qubits \cite{p1}\cite{p2} as they are ideal for state transmission. However storage of photonic qubits is difficult due to cavity leaks. Alternatively, atomic and nuclear qubits provide reliable storage \cite{a1}. The drawback with atomic and nuclear qubits is their sensitivity to interaction from their environment. These limitations affect both the teleportation protocols discussed in this paper. A shared EPR channel needs to be established either before or after Alice has the unknown qubit depending on the choice of protocol.
\\
\\
 In practical scenarios, this sharing of the EPR state is affected by interaction with the environment. Entanglement distillation is used to resolve this problem. Entanglement distillation is the process by which a number of copies of a known pure state are converted into as many Bell states as possible using LOCC, to retrieve a maximally entangled state. Additionally, the limit $2 \log_2 N$ classical bits is reasonable for ideal teleportation schemes. The total communication cost would be higher because of the LOCC operations \cite{PNE} performed during entanglement purification / distillation process .\\
\\
 For practical implementation of Kak's teleportation protocol it is imperative that a large number of copies of the unknown state be available to Alice. This inference is based on the fact that Kak protocol relies on prior availability of the unknown state before the EPR resource is shared. Chained XOR operations are performed on the unknown qubit along with an entangled qubit for eventual establishment of a quantum channel. Since entanglement distillation requires that we carry out the process recursively, more identical copies of the unknown qubits are required. This restriction could impede practical realization of Kak protocol. 

\subsection*{Conclusion}
Kak teleportation protocol halves the communication cost normally required for teleportation of an unknown qubit. Realistic implementations of this protocol might be difficult due to our inability in correcting the noise in the quantum channel using traditional means. Entanglement purification schemes which do not require a large number of unknown qubits might be more suited for practical implementations of Kak teleportation protocol. In ideal cases of standard teleportation the classical communication cost is $2 \log_2 N$. However, the total classical communication cost during realization of standard teleportation will be much higher due to factors such as entanglement purification.

\end{document}